\documentclass{article}
\usepackage{breckenr2005}
\usepackage{graphicx, epsfig}
\frompage{000} \topage{000}                                              %|
\newcommand{\pT} {\mbox{$p_T$}}
\newcommand{\pp} {\mbox{$p$$+$$p$}}
\newcommand{\pA} {\mbox{$p$$+$$A$}}
\newcommand{\NN} {\mbox{$A$$+$$A$}}

\newcommand{\xBJ} {\mbox{x$_{BJ}$}}

\newcommand{\qbar} {\mbox{$\overline{q}$}}

\def\rightmark{Future of Jets, Heavy Flavor, EM Probes}
\def\leftmark{John W. Harris}

\title{Future of Jets, Heavy Flavor, and EM Probes at RHIC and RHIC II\\}
\authors{
{John W. Harris${^1}$ %
}\\[2.812mm]
{\normalsize
\hspace*{-8pt}${^1}$ Physics Department, Yale University, \\
P.O. Box 208124, 272 Whitney Avenue, New Haven CT, U.S.A.
06520-8124}}
\abstract{Exciting results from the Relativistic Heavy Ion
Collider (RHIC) have been presented at this Workshop. However,
fundamental questions remain to be addressed in the future
regarding whether the system is deconfined, chiral symmetry is
restored, a color glass condensate exists in the initial state,
and how the system evolves through eventual hadronization. Jets,
heavy flavors and electromagnetic probes are sensitive to the
initial high density stage of RHIC collisions, and should provide
new insight. Significant additional capabilities will be added
with a luminosity upgrade of RHIC (to RHIC II), upgrades of
present detectors and a possible, new comprehensive detector at
RHIC II.}


\begin{document}

\maketitle
\setcounter{page}{1}

\section{Introduction}\label{intro}

The hard scattering and propagation of quarks and gluons,
production of heavy flavors, and electromagnetic probes are
sensitive to the initial high density stage of ultra-relativistic
heavy ion collisions, and are expected to provide important
information on the formation and properties of a Quark-Gluon
Plasma (QGP).

The hard scattering of partons (fast quarks and gluons) from the
incident nuclei occurs in the initial stage of ultra-relativistic
heavy ion collisions, within the first 1 fm/c while the nuclei
overlap. The scattered partons propagate outward and can be used
to probe subsequent stages of the collision. These partons
interact with the medium and subsequently fragment into clusters
of particles, some at large momenta, called jets. Jets are used to
determine the effects of the medium on their parent partons and
establish its properties. Properties of jets can be measured via
leading particles, particle correlations, photon-jet correlations,
heavy quark (charm or beauty) tagged-jets, and topological jet
energy. These provide information on parton energy loss,
properties of the medium through which the partons propagate,
gluon shadowing, possible existance of a color glass condensate,
and hadronization mechanisms.

The production of heavy flavor (charm and beauty) is most likely
to occur in the initial stage of the collision, since heavy quarks
require more energy to create than light quarks. Heavy flavors in
the present context include open charm, open beauty, and quarkonia
(charmonium and bottomonium states). Suppression of the yields of
different quarkonium states is predicted and depends upon the
color screening potential in a deconfined medium and the binding
strength of each individual quarkonium state. Heavy quarks that do
not result in quarkoniaum production lead to hadrons with open
charm and open beauty. Measured yields of open heavy flavors can
be used to test models for consistency of heavy flavor production
and quarkonium suppression.

Electromagnetic probes in the form of photons and leptons probe
all phases of the collision as they do not interact strongly with
the medium. These can be divided into two categories: direct
photons which tell us about thermal radiation and shadowing, and
virtual photons (electron-positron pairs) that by way of their
coupling to vector mesons may tell us about chiral symmetry
restoration and possible bound states in a strongly-coupled QGP.

%The primary physics goal of the Relativistic Heavy Ion Collider
%(RHIC) is to establish the presence of the QGP
%and determine its properties. Future physics includes: 1)
%determining the extent of deconfinement; 2) establishing a
%signature of chiral symmetry restoration; 3) possibly
%distinguishing a strongly-coupled from a weakly-interacting QGP;
%4) determining whether a color glass condensate exists in the
%initial state; and 5) understanding parton propagation and the
%hadronization process.

Many exciting results consistent with formation of the QGP have
been presented at this Workshop. Some theorists state emphatically
that evidence for the discovery of the QGP is conclusive
\cite{Gyulassy_McLerran,Wang}. Experimentalists are cautious,
stating that the evidence is not presently conclusive
\cite{STAR_white,PHENIX_white,PHOBOS_white,BRAHMS_white}. Jets,
heavy flavors and electromagnetic probes are expected to play a
key role in settling this debate, as they reflect the initial high
density stage, when formation of a QGP is most likely.

\section{RHIC Experimental Capabilities}

\subsection{RHIC Detector Capabilities}

The two large experiments at RHIC, PHENIX and STAR, continue to
improve triggering and data acquisition capabilities in order to
acquire data efficiently for jets, heavy flavors and EM probes.
PHENIX and STAR will add detector capabilities and expand
apertures with upgrades. Additional detector capabilities
\cite{STAR_plan,PHENIX_plan} include micro-vertexing for
identifying displaced vertices from heavy flavor decays in STAR
($\mu$VTX) and PHENIX (MVTX), adding better low-mass
"hadron-blind" di-lepton capabilities in PHENIX (HBD), and
extending particle identification to larger transverse momenta in
STAR (ToF) and PHENIX (Aerogel).

\subsection{RHIC Luminosities}

In order to undertake extensive studies of heavy flavor production
and jets at large transverse momenta (p$_T$), which have low cross
section (and are often called rare probes), an upgrade in the RHIC
luminosity will be necessary. The RHIC design luminosity for Au+Au
is L$_o$ = 2 $\times$ 10$^{26}$ cm$^{-2}$s$^{-1}$. RHIC now
routinely reaches twice this value. Thus, the anticipated
$\int$Ldt per RHIC year (20 weeks operation) is approximately 2 -
3 nb$^{-1}$. The p + p data are used as a reference to understand
the fundamental production mechanisms, while d + Au data provide a
comparison for the hot medium produced in Au + Au at RHIC. The p +
p reference data and the d + Au comparison/control data require
statistics similar to that of Au+Au and extended RHIC operation.

Many crucial measurements with hard and electromagnetic probes
require $\int$Ldt $>$ 20 nb$^{-1}$, which is longer than a five
year program at RHIC. For a vital program with rare probes to
continue at RHIC, a luminosity increase to 40 $\times$ L$_o$ or
$\int$Ldt $\sim$ 100 nb$^{-1}$ is planned \cite{Drees_RHIC2}. This
will be called the RHIC II Project, with a construction start
possible in 2009 for operation in 2012.

\section{Jets}

High transverse momentum (p$_T$) particles and jets can be used to
probe the QGP, study its properties and gain a better
understanding of high density QCD and hadronization
\cite{Bellwied}. Measuring the modifications of fragmentation
functions (FF) of partons traversing the QGP in A+A collisions
relative to p+p and p+A collisions should identify the properties
of the QGP compared to those of a nuclear or hadronic medium. The
RHIC energy regime appears to be ideal for these studies. Recent
measurements in the forward direction at RHIC indicate possible
gluon shadowing in the initial state at low x. Therefore,
measurements over a specific part of phase space (e.g. forward- or
mid-rapidities) selects the x region of the dominant process of
interest. The higher energy regime of the Large Hadron Collider
(LHC) will provide increased particle yields at high p$_T$  and at
low x.

The contributions of the various (u, d, s, c, b) quarks to the
mass of stable particles can be extracted by measuring the
fragmentation function of each particle in $\pp$ interactions.
Bourrely and Soffer \cite{Bourrely-Soffer} use a statistical model
for the fragmentation functions of the octet baryons (p,
$\lambda$, $\Sigma$, and $\Xi$) in e$^{+}$e$^{-}$ interactions at
Q = 91.2 GeV. They find that the contributions of the light
(u,d,), strange (s), and heavy (c,b) quarks to the production of
these particles varies as a function of $\xBJ$, with the
fragmentation of heavy quarks dominating the fragmentation protons
for parton momentum fractions x $\leq$ 0.3 (see \cite{Bellwied}
for more details). In order to measure these fragmentation
functions at RHIC, leading particles in jets with large transverse
momentum must be identified. Such measurements in $\NN$ collisions
will establish how fragmentation functions are modified by
propagation of the various types of quarks in the dense medium and
will reflect the various quark contributions to the particle
masses as they fragment in the medium. It would be extremely
exciting if fragmentation functions of some of the particles were
to reflect properties of a chirally restored medium. In addition
to accounting for the constituent quark masses, the chiral quark
condensate is responsible for inducing transitions between
left-handed and right-handed quarks, $\qbar q$ = $\qbar_{L} q_{R}$
+ $\qbar_{R} q_{L}$. Therefore, helicities of leading particles in
jets (determined from the polarization of leading $\Lambda$
particles) may provide information on parity violation and chiral
symmetry restoration \cite{Kharzeev-Sandweiss}.

Full utilization of hard parton scattering to probe high density
QCD matter requires measurements of photons (to establish the
parton energy), jets (another potential measurement of parton
energy), high p$_T$ identified particles (for fragmentation
functions of particles and flavor-tagging), as well as intrajet,
jet-jet and photon-jet correlations. In addition, measurements are
essential over a multi-parameter space that can be divided into
initial state parameters (c.m. energy, system mass, collision
impact parameter, x$_1$ and x$_2$ of the colliding beam partons,
and Q$^2$ of the collision) and those of the final state
(p$_T^{parton}$, y$^{parton}$, $\phi^{parton}$,
p$_T^{jet/particle}$, y$^{jet/particle}$, $\phi^{jet/particle}$,
flavor$^{jet/particle}$, $\phi^{flow plane}$). Such a
comprehensive study requires large data sets and high luminosity
to extend measurements to large p$_T$. Furthermore, it would be of
interest to investigate the difference in quark versus gluon
propagation by implementing kinematic cuts (x$_1$, x$_2$) in
jet-jet correlations, and to utilize the expected differences
predicted by QCD for the yields of gluon and quark jets as a
function of transverse momentum and $\sqrt{s}$.

\subsection{Photon-tagged Jets}

The primary advantage of photons is that they do not re-interact
with the medium through which they propagate. Thus, they can be
used to determine the parton energy in the original
hard-scattering that produces the photon and away-side jet. The
away-side jet will suffer energy loss in the medium and thus the
difference of the photon and measured away-side jet energy can be
used to determine the energy lost by the parton on the away-side
of the photon. Correlation of the flavor of the leading hadron in
the jet on the away-side of a photon provides additional
information with which to test energy loss mechanisms. Another
advantage of photon-jet correlations is that there are only two
production diagrams that produce photons in leading order:
quark-antiquark annihilation and quark-gluon Compton scattering.
However, there is one major issue to be dealt with when utilizing
prompt photons to determine the momentum of the hard-scattered
parton. One must distinguish direct (prompt) photons from
fragmentation photons. This is accomplished by using isolation
cuts in elementary interactions. However, it is complicated by the
large particle multiplicities in A + A collisions, and must be
resolved through understanding the contribution of fragmentation
photons and their momentum dependence.

STAR will undertake initial studies of photon-jet correlations up
to 10 GeV/c photon momentum by utilizing a few nb$^{-1}$ integral
luminosity in Au+Au at RHIC. Approximately 1 percent of the jets
have a leading hadron above background in a Au+Au collision at
RHIC. A few year Au+Au run with 4 - 5 nb$^{-1}$ integral
luminosity yields $\sim$ 8K charged hadrons in a spectrum on the
away-side from a 10 GeV/c photon, and $\sim$ 1K charged hadrons in
a spectrum on the away-side from a 15 GeV/c photon in STAR. More
detailed measurements, especially with identified particles on the
away-side for fragmentation function modification of partons
requires RHIC II luminosities and additional particle
identification at large p$_T$.

PHENIX proposes to undertake statistical photon-jet correlation
analyses with approximately 1000 photon-jet events. The maximum
photon p$_T$ depends on the PHENIX detector complement. In the
2004 Au+Au run the maximum photon p$_T$ (p$_T^{max}(\gamma)$) is
expected to be $\sim$ 6 GeV/c; with a new time projection chamber
(covering $-1 \leq \eta \leq 1$) at RHIC luminosity
p$_T^{max}(\gamma) \sim$ 12 GeV/c; with an additional nose cone
calorimeter for expanded photon detection at RHIC II luminosity
p$_T^{max}(\gamma) \sim$ 23 GeV/c at mid-rapidity \cite{Drees}.

\begin{figure}[htb]
\vspace{-\baselineskip}
  \hspace{\fill}
    \begin{center}
    \includegraphics[width=0.5\textwidth]{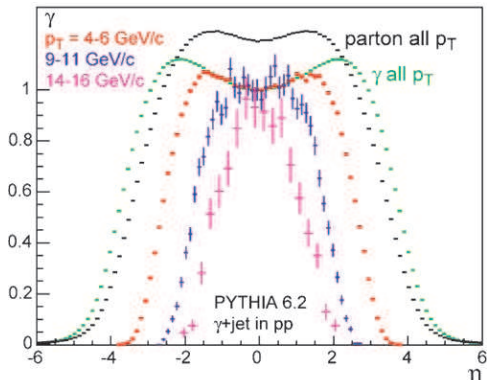}
    \vspace{-\baselineskip}
    \caption{\small{Photon-jet results from PYTHIA 6.2
    for $\sqrt{s}$ = 200 GeV p+p interactions. Distributions as a
    function of pseudo-rapidity $\eta$ are shown from
    top to bottom for partons of all \pT, photons of all \pT, and
    photons with momenta
    4 $<$ \pT $<$ 6 GeV/c, 9 $<$ \pT $<$ 11 GeV/c, and 14 $<$ \pT $<$ 16
    GeV/c, respectively.}}
    \label{fig_jets1}
    \end{center}
\vspace{-1\baselineskip}
\end{figure}

\begin{figure}[htb]
\vspace{.5\baselineskip}
  \hspace{\fill}
    \begin{center}
    \includegraphics[width=0.7\textwidth]{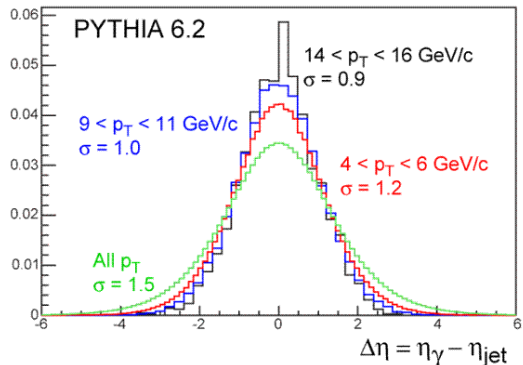}
    \vspace{-3\baselineskip}
    \caption{\small{Photon-jet results from PYTHIA 6.2
    for $\sqrt{s}$ = 200 GeV p+p interactions. Distributions for the difference in
    pseudo-rapidity between a direct photon and its
    associated away-side jet
    for parton or direct photon momenta ranging (top to bottom) from
    14 $<$ \pT $<$ 16 GeV/c, 9 $<$ \pT $<$ 11 GeV/c,
    4 $<$ \pT $<$ 6 GeV/c, and all \pT, respectively. The standard deviation
    $\sigma$ of each distribution is also given.}}
    \label{fig_jets2}
    \end{center}
\vspace{-2\baselineskip}
\end{figure}

Recent results from STAR \cite{Magestro} indicate that jets in
Au+Au collisions broaden significantly in pseudo-rapidity on the
near-side as well as on the away-side. This can be attributed both
to the variation of parton momentum fractions for partons of the
incoming beam nuclei (parton momentum fractions x$_1$,x$_2$) and
to jet quenching in matter. The broadening in p+p interactions can
be seen in calculations utilizing PYTHIA 6.2 for $\sqrt{s}$ = 200
GeV at RHIC, shown in Fig. \ref{fig_jets1}. The partons and
photons from hard scattering processes extend widely over the
pseudo-rapidity range $-3 \leq \eta \leq 3$. The same simulations
show that the difference in pseudo-rapidity between a direct
photon and associated away-side jet, shown in Fig.
\ref{fig_jets2}, has a width $\sigma$($\eta_{\gamma} -
\eta_{jet}$) = 0.9, 1.0, 1.2, and 1.5 units of pseudo-rapidity for
parton or direct photon momenta ranging from 14 $<$ \pT $<$ 16
GeV/c, 9 $<$ \pT $<$ 11 GeV/c, 4 $<$ \pT $<$ 6 GeV/c, and all \pT,
respectively. Experiments seeking to undertake photon-jet
measurements should have large acceptance for photons, high p$_T$
particles and jet energy in order to cover a range of parton
momentum fractions (x$_1$,x$_2$).

\subsection{Flavor-tagged Jets}

Jets with a D- or B-meson as leading particles serve to study the
response of the medium to heavy compared to light quarks. A high
p$_T$ electron in coincidence with a leading hadron, both from a
vertex displaced from the primary reaction vertex, will provide a
trigger for heavy flavor decays. Significant measurements at large
p$_T \sim$ 10 - 15 GeV of D- and B-mesons as leading particles of
jets require RHIC II luminosities and upgraded detectors in STAR
($\mu$-vertex, ToF) and PHENIX (VTX).\textbf{\textbf{}}

\section{Heavy Flavors}

\subsection{Quarkonia}

The production of quarkonium states in p+p, p+A, and A+A
collisions provides a tool to study deconfinement in strongly
interacting matter \cite{Matsui}. Studies of the dependence of the
heavy-quark potential on the in-medium temperature in lattice QCD
calculations with dynamical quarks \cite{Digal} indicate a
sequence of melting of the quarkonium states based upon their
binding strengths: T($\psi$') $<$ T($\Upsilon_{3S}$) $<$
T(J/$\psi$) $\sim$ T($\Upsilon_{2S}$) $<$ T($\Upsilon_{1S}$) where
T($\Upsilon_{3S}$) $<$ T$_c$ and T($\Upsilon_{1S}$) $>$ T$_c$,
with T$_c$ the deconfinement phase transition temperature.
Therefore, a measurement of the yields of the various bottomonium
states will shed light on the production (via $\Upsilon_{1S}$) and
suppression mechanisms ($\Upsilon_{2S}$ and $\Upsilon_{3S}$) of
quarkonia avoiding many difficulties inherent in charmonium
measurements. These measurements are challenging, requiring
excellent momentum resolution to resolve the bottomonium states
and very high rate (luminosity) and trigger capabilities because
of the low production cross-sections.

The larger production cross-sections for charmonium states
compared to bottomonium states have led to studies of charmonium
and the subsequent observation of charmonium suppression in
collisions of heavy ions at the SPS. PHENIX anticipates its first
results with large statistics on charmonium suppression in Au+Au
at RHIC from the large statistics 2004 data run. Bottomonium
spectroscopy on the other hand requires higher luminosities. Since
bottomonium is massive ($\sim$ 10 GeV/c$^2$) its decay leptons
have sufficiently large momenta above background processes
facilitating high-level triggering.

The PHENIX mass resolution for $\Upsilon$ $\rightarrow e^{+}e^{-}$
with the VTX detector upgrade is $\Delta$m = 60 MeV. Without the
VTX it is 170 MeV, making resolution of the $\Upsilon_{1S}$ (9.460
GeV), $\Upsilon_{2S}$ (10.020 GeV), and $\Upsilon_{3S}$ (10.360
GeV) challenging. The PHENIX mass resolution in the muon arms is
worse than 170 MeV making it difficult to resolve the individual
$\Upsilon$ states in the muon decay channel. Statistics for the
$\Upsilon$ states, combined in Table 1, are low. In STAR, the mass
resolution for $\Upsilon$ $\rightarrow e^{+}e^{-}$ is $\Delta$m =
340 MeV using the time projection chamber tracking alone. A
$\mu$-vertex detector upgrade would improve this resolution to
$\Delta$m = 170 MeV. Only with a planned data acquisition system
upgrade, will STAR be able to detect a significant number of
$\Upsilon$'s (1750 combined in all three states with 1.5 nb$^{-1}$
Au+Au). In general, a meaningful bottomonium program at RHIC will
require RHIC II luminosities ($\sim$ 100 nb$^{-1}$) and large
acceptances to obtain reasonable statistics.

The quarkonium statistics anticipated for Au + Au in PHENIX
\cite{PHENIX_plan,Drees} are presented in Table 1.

\begin{table}[hbt]
\vspace {-.3cm} \centering \caption{PHENIX Quarkonium Program for
Au+Au at RHIC and RHIC II. \cite{PHENIX_plan,Drees}} \label{tab:1}
\begin{tabular}{lll}
%\hline\noalign{\smallskip}
Channel   & RHIC (1.5 nb$^{-1}$)    & RHIC II (30 nb$^{-1}$) \\
\noalign{\smallskip}\hline\noalign{\smallskip}
J/$\psi \rightarrow$ e$^+$e$^-$ & 2,800 & 56,000 \\
$\psi$' $\rightarrow$ e$^+$e$^-$ & 100 & 2,000 \\
$\Upsilon$ $\rightarrow$ e$^{+}$e$^{-}$ & 8$\dag$ & 155$\dag$\\
(all states)& & \\
   &  &  \\
J/$\psi$ ($\psi$') $\rightarrow \mu^{+} \mu ^{-}$ & 38,000 (1400) & 760,000 (28,000)\\
$\Upsilon$ $\rightarrow \mu^{+} \mu ^{-}$  & 35$\ddag$ & 700$\ddag$\\
(all states)& & \\
%\noalign{\smallskip}\hline \\
\end{tabular}
\end{table}
\vspace {-.3cm}\hspace{-.75cm}
$\dag$ requires MVTX upgrade\\
$\ddag$ requires $\mu$-trigger system upgrade.\\

\vspace{-\baselineskip}
\subsection{Open Heavy Flavor}

Heavy flavor (charm and bottom) yields are sensitive to the
initial gluon density and are important components of
understanding J/$\psi$ and $\Upsilon$ production. Measuring the
energy loss of a heavy quark in the medium will indicate whether
heavy quarks suffer less energy loss than light quarks in the
medium as expected from the "dead-cone effect" \cite{Dokshitzer}.
Also, the stronger quenching of gluons than quarks results in
stronger quenching (and a stronger energy dependence of quenching)
for light than heavy mesons, due to the contribution of gluons to
light meson production \cite{Djordjevic}.

Initial measurements of charm cross sections and charm flow have
been made by identifying single electrons above background in STAR
\cite{STAR_electrons} and PHENIX \cite{PHENIX_electrons}. These
results indicate that low p$_T$ open charm exhibits elliptic flow
and are preliminary at the time of this Conference. Significant
measurements of charm flow and charm jet energy loss up to
moderate p$_T \sim$ 5 - 6 GeV can be made with $\sim$ 3 nb$^{-1}$
and upgraded detectors in STAR ($\mu$-vertex, ToF) and PHENIX
(VTX). Recent calculations of the color charge and mass dependence
of the energy loss indicate that the p$_T$ range for identifying
significant differences in jet quenching of heavy versus light
mesons at RHIC is 7 $<$ p$_T$ $<$ 12 GeV/c \cite{Armesto}. Such
measurements require the increased machine and detector
capabilities of RHIC II.

\section{Electromagnetic Probes}

\subsection{Direct Photons}

Photons in A+A collisions may provide information on thermal
photon radiation. In p+A interactions, photons establish the
degree of shadowing. Photons in p+p reactions are needed for
reference data, to understand the underlying processes in p+A and
A+A results. Preliminary results on photons in Au+Au and p+p at
RHIC have been reported by PHENIX at this Workshop
\cite{Sakaguchi}. Photons measured in p+p are consistent with
next-to-leading order (NLO) pQCD calculations. Those measured in
Au+Au exhibit no thermal photons within present statistics.
Furthermore, the Au+Au direct photons are consistent with binary
scaling of p+p. A definite statement about direct photons in Au+Au
at RHIC is anticipated from PHENIX from the recent RHIC 2004 high
statistics Au + Au run \cite{Sakaguchi}.

\subsection{Virtual Photons via e$^+$e$^-$ pairs}

Thermal photons (measured via e$^+$e$^-$ pairs) are expected to be
radiated from the QGP.  However, at RHIC energies, the thermal
di-lepton spectrum in the intermediate mass range (1 - 3 GeV) may
be dominated by charm. In addition to information on thermal
radiation, virtual photons (measured via e$^+$e$^-$ pairs)
investigate possible modifications of vector mesons in the medium.
The behavior of vector mesons in medium may shed light on the
existence of chiral symmetry breaking and/or bound states in a
strongly-coupled QGP.

\begin{figure}[htb]
\vspace{-\baselineskip}
  \hspace{\fill}
    \begin{center}
    \includegraphics[width=0.6\textwidth]{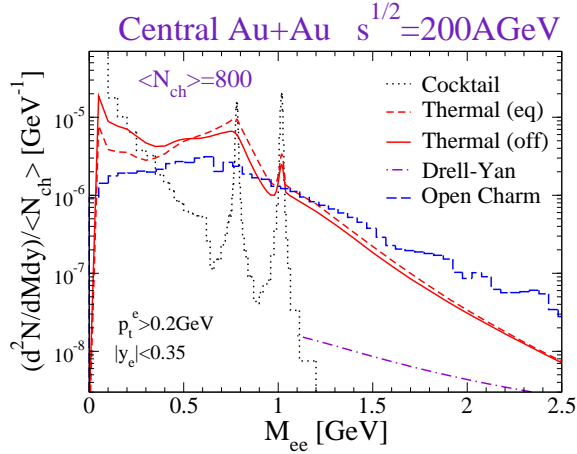}
      \caption{\small{Low mass e$^+$e$^-$ pair spectrum
      from \cite{Rapp}. Contributions from the
      hadronic "cocktail", thermal and non-equilibrium photons,
      Drell-Yan, and open charm are displayed as denoted in the
      figure.}}
      \label{fig1}
      \vspace{-\baselineskip}
    \end{center}
    \vspace{-\baselineskip}
\end{figure}

In order to effectively pursue low mass electron pair
measurements, PHENIX has proposed to install a hadron-blind TPC
(HBD) and STAR has proposed a barrel Time-of-Flight detector (ToF)
for electron identification at p$_T$ $>$ 0.2 GeV/c. A calculation
of the light vector meson yield as a function of invariant mass of
e$^+$e$^-$ pairs is displayed in Fig. \ref{fig1} \cite{Rapp}.
Peaks for the $\omega$ and $\phi$ mesons are observed, but are
swamped by e$^+$e$^-$ pairs from thermal and non-equilibrium
photons and open charm. Careful investigation of medium
modification of low mass vector mesons requires measuring and
understanding all contributions to the di-lepton spectrum
including detailed charm studies that require RHIC II
luminosities.

\section{A Comprehensive New RHIC II Detector}

\begin{figure}[htb]
\vspace{-3\baselineskip}
  \hspace{\fill}
    \begin{center}
    \includegraphics[width=0.75\textwidth]{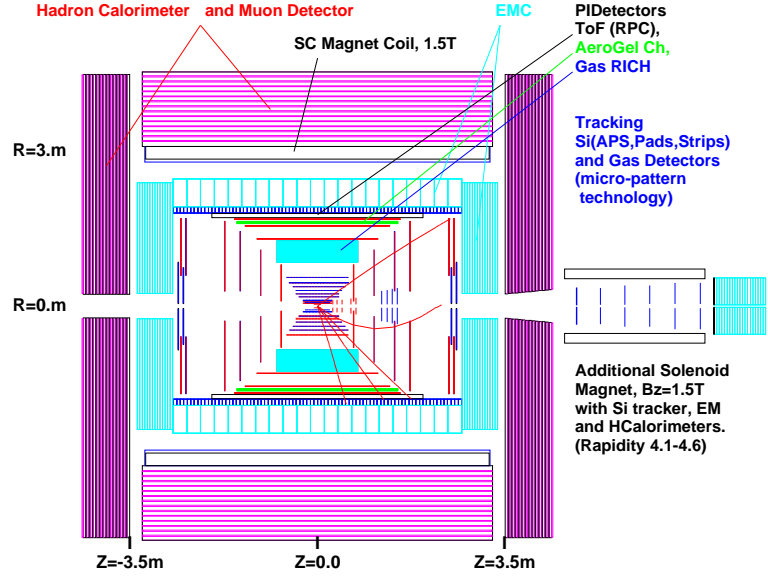}
      \caption{\small{Diagram of a possible, comprehensive new detector
      at RHIC II using the SLD magnet.}}
      \label{fig3}
    \end{center}
\vspace{-2\baselineskip}
\end{figure}

A comprehensive new detector has been proposed
\cite{new_detector_EoI} for RHIC II to undertake measurements of
jets, heavy flavors, and electromagnetic probes, and to take full
advantage of the high RHIC II luminosities. New RHIC II physics
opportunities can be studied by utilizing a high field ($\sim$ 1.5
T) magnet, extensive charged hadron tracking and identification at
high p$_T$, electron and muon tracking and identification, and
extensive coverage of electromagnetic and hadronic calorimetry.
The capabilities of a comprehensive new detector include: 1)
excellent charged particle momentum resolution to p$_T$ = 40 GeV/c
in the central rapidity region, 2) complete hadronic and
electromagnetic calorimetry over a large phase space (-3 $\leq$
$\eta$ $\leq$ 3, $\Delta \phi$ = 2$\pi$), 3) particle
identification out to large p$_T$ ($\sim$ 20 - 30 GeV/c) including
hadron ($\pi$, K, p) and lepton (e/h, $\pi$/h) separation in the
central and forward regions, and 4) high rate detectors, data
acquisition, and trigger capabilities. A possible layout for a new
RHIC II detector using the SLD magnet is shown in Fig. \ref{fig3}.

In order to identify all charged hadrons in a high $\pT$ jet at
RHIC II, hadron identification is necessary up to momenta of
approximately 20 GeV/c. Lepton particle identification will be
achieved through the e/h capabilities in the calorimeters and the
muon chambers. Hadron and lepton particle identification will be
achieved through a combination of dE/dx in the tracking detectors
($p_T$ $\leq$ 1 GeV/c), a time-of-flight device ($p_T$  $\leq$ 3
GeV/c), and a combination of two different Aerogel
Cherenkov-threshold counters and a RICH detector with gas radiator
(up to $p_T$ $\sim$ 20 GeV/c). For more details on the
comprehensive new detector see \cite{new_detector_EoI}.

\subsection{Jets in a Comprehensive New Detector}

Jet rates over the large acceptance of a new detector (-3 $\leq$
$\eta$ $\leq$ 3, $\Delta \phi$ = 2$\pi$ plus extended forward
coverage to $\eta \sim$ 4.5) at upgraded RHIC II luminosities will
be significant. The anticipated jet yield for 40 GeV jets in such
a new detector at RHIC II with 30 nb$^{-1}$ of Au+Au at top energy
is $\sim$ 180,000. 19,000 $\gamma$-jet events are expected with
p$_{T}(\gamma)$ = 20 GeV/c, and 1,000 $\gamma$-jet events for
p$_{T}(\gamma)$ = 30 GeV/c with full away-side particle
identification over -3 $\leq$ $\eta$ $\leq$ 3 for determination of
the modification of fragmentation functions of particles.

Extension of high resolution particle tracking, particle
identification (PID), and calorimetry to forward rapidities will
be important in elucidating the various particle production and
hadronization mechanisms, which may be sensitive to the quark and
gluon components of the hadronic wave functions. At very low
$\xBJ$, gluons may be coherent over nuclear distances forming a
color glass condensate \cite{McLerran,Kharzeev-Levin}. This would
have effects on many hard physics observables that depend directly
on the gluon structure, e.g. minijet rates and heavy flavor
production, but can be clarified by comparisons of $\pA$ physics
with $\pp$. Thus, it is important to study high $\pT$ processes
away from midrapidity. To take full advantage of physics in the
forward region, momentum measurements and PID must be undertaken
up to $\pT$ $\sim$ 2-3 GeV/c, which is a real experimental
challenge with longitudinal momenta of 20-30 GeV/c at large
rapidities.

\subsection{Quarkonia in a Comprehensive New Detector}

For determination of the quarkonium melting sequence an energy
resolution of better than 10$\%$/$\sqrt(E)$ is required to resolve
the quarkonium states with calorimeter information alone. Thus,
quarkonium physics at RHIC II in this new detector will fully
utilize an electromagnetic calorimeter in combination with high
resolution tracking and large acceptance muon chambers. The mass
resolution for $\Upsilon$ $\rightarrow \mu^{+} \mu^{-}$ in the new
comprehensive detector is $\Delta$m = 60 MeV. Furthermore, large
acceptance in the Feynman x$_F$ variable is important for
understanding quarkonium production and melting mechanisms. This
leads to the need for large acceptance in $\eta$ for lepton pairs.
The electron and muon coverage of the new detector extends over -3
$\leq$ $\eta$ $\leq$ 3 and $\Delta \phi$ = 2$\pi$. A similar
acceptance in the new detector for charmonium feed-down photons
from $\chi_c$ decays ($\chi_{c} \rightarrow$ J/$\psi$ + $\gamma$)
allows determination of the $\chi_c$ feed-down contribution to
J/$\psi$ production and subsequent suppression.

Anticipated quarkonium statistics in Au + Au for the comprehensive
new detector are presented in Table 2, which can be compared
directly to Table 1.

\vspace{-\baselineskip}
\begin{table}[h]
\centering \caption{Quarkonium Program in Au + Au for a
Comprehensive New Detector at RHIC II for p$_{lepton}$ $>$ 2 GeV/c
for J/$\psi$, and p$_{lepton}$ $>$ 4 GeV/c for $\Upsilon$.}
\label{tab:1}
\begin{tabular}{lll}
\hline\noalign{\smallskip}
Channel   & RHIC II (30 nb$^{-1}$) \\
\noalign{\smallskip}\hline\noalign{\smallskip}
J/$\psi \rightarrow$ di-leptons & 36,000,000 \\
$\psi$' $\rightarrow$ di-leptons & 1,000,000 \\
$\chi_c$' $\rightarrow$ J/$\psi$ + $\gamma$ & 680,000 \\
$\Upsilon$ $\rightarrow$ di-leptons & 64,000\\
$\Upsilon$' $\rightarrow$ di-leptons  & 12,000\\
$\Upsilon$'' $\rightarrow$ di-leptons  & 12,000\\
\noalign{\smallskip}\hline \\
\textbf{}
\end{tabular}
\vspace{-2\baselineskip}
\end{table}

\vspace{-\baselineskip}
\section{Conclusions}

There is new data still to be accumulated at RHIC utilizing jets,
heavy flavors and electromagnetic probes. From this data new
physics will be uncovered, since jets, heavy flavors and
electromagnetic probes are sensitive to the initial high density
stage of RHIC collisions. Questions still remain to be addressed
as to whether 1) the system becomes deconfined, 2) chiral symmetry
is restored, 3) in addition to a strongly-coupled QGP there is a
weakly-interacting one, 4) a color glass condensate exists in the
initial state, and 5) whether we can gain new understanding of the
hadronization process. Precise timescales for new detector
implementation to improve capabilities for rare probes at RHIC is
uncertain due to ambiguities in the availability of funding.
Significant capabilities will be added with new detectors at RHIC
and a possible comprehensive new detector at RHIC II.

\section{Acknowledgements}

The author wishes to thank R. Bellwied, T. Ullrich, N. Smirnov, P.
Steinberg, H. Caines, M. Lamont, C. Markert, J. Sandweiss, M. Lisa
and D. Magestro for fruitful RHIC II physics discussions and
contributions to this work. M. Gyulassy, B. Mueller and D.
Kharzeev have contributed through enlightening discussions.

\vfill\eject
\end{document}